\documentclass[pre,twocolumn,showpacs,aps]{revtex4-1}

\usepackage{amssymb,amsfonts,amsmath}
\usepackage[pdftex]{graphicx}
\usepackage{multirow}
\usepackage{color}

\begin{document}

\title{Influence of opinion dynamics on the evolution of games}

\author{Floriana Gargiulo}\affiliation{Centre d'Analyse et de Math\'ematique Sociales CAMS, CNRS, 190 Avenue de France, 75013 Paris, France}
\author{Jos\'e J. Ramasco}\affiliation{Instituto de F\'{\i}sica Interdisciplinar y Sistemas Complejos IFISC (CSIC-UIB),\\ 07122 Palma de Mallorca, Spain}

\widetext

\begin{abstract} 
Under certain circumstances such as lack of information or bounded rationality, human players can take decisions on which strategy to choose in a game on the basis of simple opinions. These opinions can be modified after each round by observing own or others payoff results but can be also modified after interchanging impressions with other players. In this way, the update of the strategies can become a question that goes beyond simple evolutionary rules based on fitness and become a social issue. In this work, we explore this scenario by coupling a game with an opinion dynamics model. The opinion is represented by a continuous variable that corresponds to the certainty of the agents respect to which strategy is best. The opinions transform into actions by making the selection of an strategy a stochastic event with a probability regulated by the opinion. A certain regard for the previous round payoff is included but the main update rules of the opinion are given by a model inspired in social interchanges. We find that the dynamics fixed points of the coupled model is different from those of the evolutionary game or the opinion models alone. Furthermore, new features emerge such as the independence of the fraction of cooperators with respect to the topology of the social interaction network or the presence of a small fraction of extremist players.
\end{abstract}

\maketitle 

\section{Introduction}

Evolutionary game theory has been introduced as a framework to study the processes of selection of genes or behaviors in biological and social systems~\cite{maynard73,maynard82,nowak06}. Its aim is to characterize the choices in terms of strategies of individuals of a population playing a game. A particular strategy generates a payoff to the individual playing it that depends on the selection of the rest of individuals. The key assumption of the evolutionary theory is that the fitness of an individual to reproduce directly relates to the payoff obtained~\cite{maynard73}. Consequently, most successful strategies in terms of payoff are also those that multiply faster and can eventually become dominant after some generations. 

These ideas find an analytical expression in the form of the so-called replicator equation~\cite{schuster83,nowak04,schlag98}. If $x_i$ stands for the fraction of individuals in the population playing strategy $i$, $f_i(\vec{x})$ for their payoff and $\bar{f}(\vec{x})$ for the average payoff over all the population, the replicator equation reads
\begin{equation}
\frac{d x_i}{dt} = x_i \, \left( f_i(\vec{x}) - \bar{f}(\vec{x}) \right) .
\label{rep}
\end{equation}
The fixed points and limit cycles of the equation define the final state of the system regarding the distribution of strategies in the population~\cite{nowak06,schuster83,nowak04,taylor78}. Moreover, the study of the stability of the solutions, particularly if they are formed by single strategies, to invasion by other strategies motivates the definition of evolutionary stables strategies (ESS)~\cite{taylor78}. To illustrate the predictions of this approach, one can consider the social dilemmas such as the public goods game or the prisoner's dilemma. In these games, each individual must choose between collaborating with her partners getting a intermediate value of the payoff or to defect and try to take advantage of those partners that are collaborating to gain a higher payoff. Despite collaboration is beneficial to the population as a whole, the egoist inclination of each single individual to maximize her payoff leads to generalized defection as the replicator equation predicts since this is the only stable solution ~\cite{nowak06,schuster83}.

This result can seem a little drastic especially when considered in the light of everyday experience in human societies or the known behavior of social animals. Different mechanisms have been proposed to explain how the collaboration levels can increase in a population.  One is, for instance, taking into account the finite and discrete character of the individuals in the population. This point goes beyond the assumptions of the continuous theory and provide thus a escape door to obtain more collaboration or even to the invasion of collaborative individuals in a full-defect population~\cite{nowak04b,claussen08,vilone11}. However, its efficiency as an explanation does not extend to large systems since the probability of survival or invasion of collaborative strategies decreases fast with the population size. Other possibility that has been theoretically discussed is that structured populations may increas collaboration. Geographical extended systems simulated using spatial lattices show a remanent level of collaboration~\cite{nowak92,blume93,sysi05,roca09} and even chaotic patterns separating areas of collaborating and defecting individuals~\cite{nowak92}. The structure of social networks enhances collaboration via the heterogeneity of individual roles that the different positions in the network produce~\cite{zimmermann04,santos06,gomez07,szabo07,santos08}. Also random mutations or the individuals' free exploration to search for a best response to the strategies of their counterparts are another element that can promote collaboration~\cite{sysi05,foster90,blume99,szabo05,traulsen09}. Finally, the fixed points of the system dynamics, including the level of cooperations, are affected too by the way in which the system updates either by taking into account discrete versus continuous dynamics~\cite{benaim02,roca06} or by altering the update rules~\cite{hofbauer03,traulsen05,vilone12}.  

\begin{figure}
\begin{center}
\includegraphics[width=8.6cm]{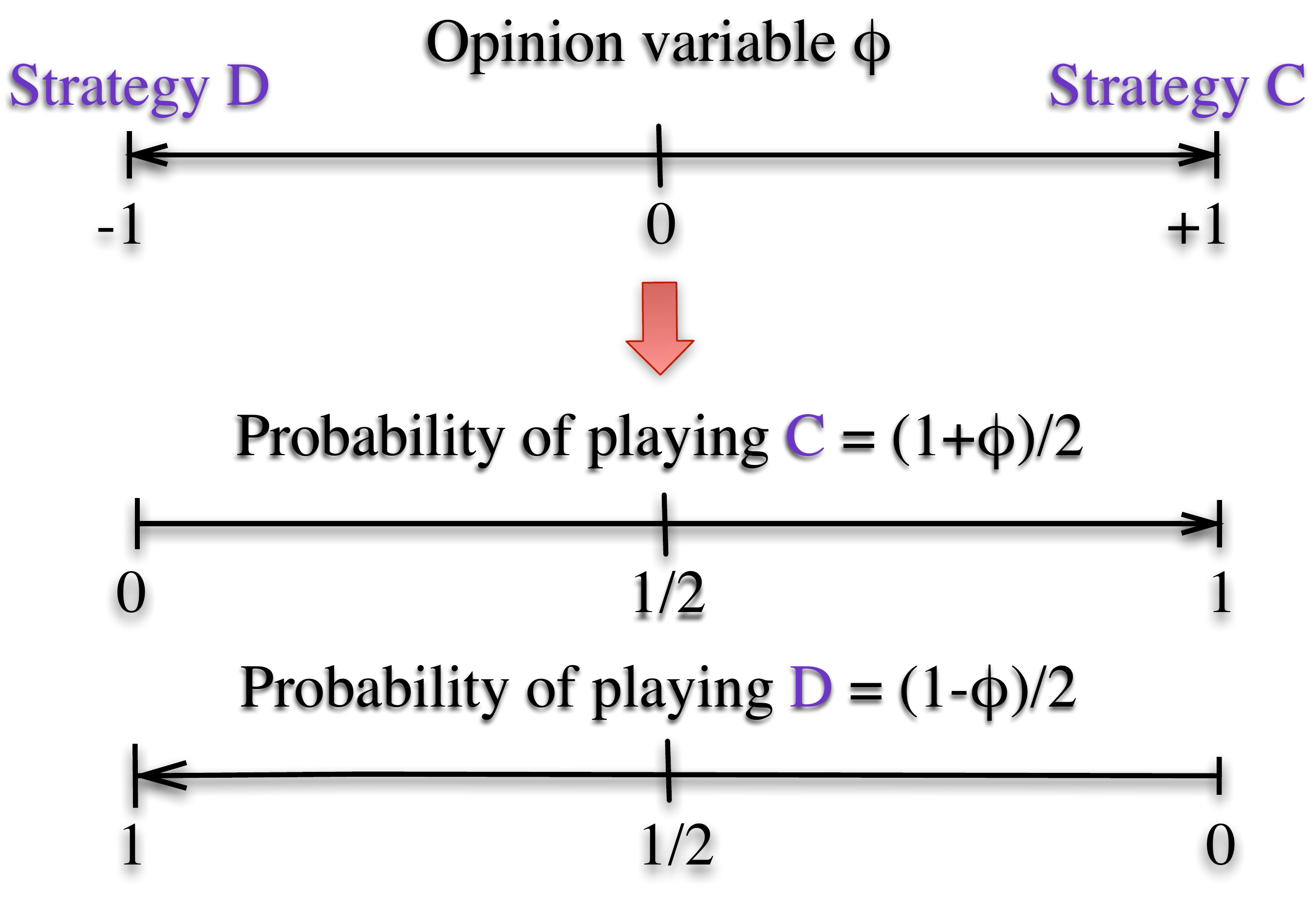}
\caption{Sketch showing the coupling between the opinion variable $\phi$ and the probability of opting for one of the two strategies in the game collaboration (C) or defection (D).\label{sketch}}
\end{center}
\end{figure}

\begin{figure*}
\begin{center}
\includegraphics[width=18cm]{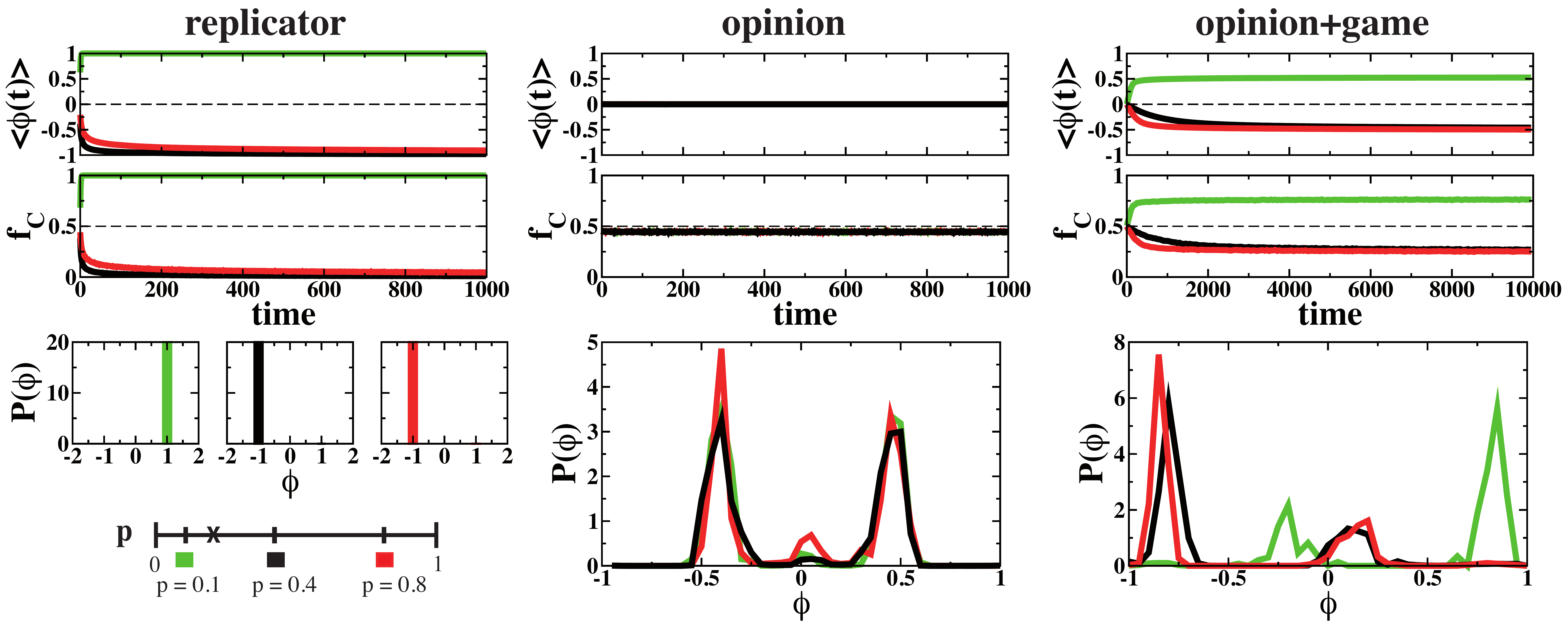}
\caption{Time evolution of the average opinion and of the fraction of agents playing collaborate ($f_C$) for the replicator dynamics, the opinion dynamics alone and the coupled dynamics of game and opinion. The results are shown for three different values of the parameter $p$: $0.1$, $0.4$ and $0.8$. Note that $C$ is the most advantageous strategy for $p < 0.2$, while the game becomes a dilemma for $p > 0.2$. The bottom plots show the probability density for the opinion of the agents at the last time of simulation. The simulations are run with a value of $\epsilon = 1/2$ and a population of $N = 5000$ agents.\label{comp}}
\end{center}
\end{figure*}

In this work, we explore a mechanism that can also play an important role to raise collaboration levels in social systems. The basic idea goes back to the fact that humans not always take the most rational option when presented with a dilemma~\cite{peyton93,gardner09,delton11}. This has been observed in experiments in controlled  environments in which participants, in general students, were playing Prisoner's dilemma~\cite{andreoni93,traulsen10,grujic10,suri10,gracia12,gracia12b}. Also, in other level, it is a well known behavior in the world of finances where decisions on buying and selling can be taken based on rumors or on a general state of opinion about the possibilities of an investment~\cite{cristelli11}. Our proposal is to increase the dimensionality of the system by noting that the opinion on which is the best strategy is an important variable to incorporate, even though in some cases such belief can be wrong or baseless with respect to actual performance in the game. The evolution of the system includes thus a purely social ingredient related to opinion formation~\cite{castellano09} followed by a process of decision taking that relies on the formed opinion. In the abstract representation of Equation~(\ref{rep}), the addition of a variable of opinion can be modeled  as
\begin{equation}
\begin{array}{l}
\frac{d x_i}{dt} = g(\phi_j,x_i)  , \\
$\,$\\
\frac{d\phi_j}{dt} = h(\vec{\phi},\vec{x}) ,
\end{array}
\end{equation}
where the index describing the opinion $j$ can be continuous or, as in this example, discrete, $\phi_j$ represents the fractions of individuals holding opinion $j$, $g()$ is a function that relates the opinion $j$ with the probability of playing strategy $i$ and the function $h()$ describes the evolution of the opinions given the state of the system and the outcome of the game. The addition of the new field $\phi$ corresponding to the opinions of the individuals and the new rules of update given by the interchange of opinions between individuals can lead to extremely different fixed points and solutions for this system. In the following, we provide an example with a simple model that shows how these ideas can be implemented in practice and how the dynamic and stationary predictions of evolutionary game theory can dramatically change due to the coupling between opinion and games.

\section{Model}

We take as basis a well-known model for the opinion dynamics, the Deffuant model~\cite{deffuant00}, and a game inspired by the dilemma of the tragedy of commons~\cite{hardin68,hardin95}. The opinion in the Deffuant model is described by a continuous variable $\phi$ between $-1$ and $1$. Considering a population of $N$ agents, each one placed on a node of a network, the update of opinions is carried out by randomly choosing an agent $i$ and one of her neighbors $j$ and comparing their opinions at time $t$, $\phi_i$ and $\phi_j$. If $|\phi_i-\phi_j| < \epsilon$, the interaction occurs and the new opinions are given by 
\begin{equation}
\begin{array}{c}
\phi_i(t+1) = \phi_i + \mu \,(\phi_j - \phi_i), \\
\phi_j(t+1) = \phi_j + \mu \,(\phi_i - \phi_j).
\end{array}
\label{deff}
\end{equation}
Otherwise, if the difference between $\phi_i$ and $\phi_j$ is larger than $\epsilon$, no interaction happens. The parameter $\mu$ is the so-called convergence parameter since it regulates to which new value the opinions converge after interaction. In this work, we set it at $\mu = 1/2$ which implies that the final opinion is the average over both agents opinion. It is important to stress that Deffuant's model shows bounded confidence in the sense that interactions between agents whose opinions are further apart than $\epsilon$ are forbidden. The value of $\epsilon$ is thus a key parameter to take into account in the following study.

For the game, we consider a simple set of rules that permit the exploration  of a dilemma and a harmony scenario by tuning a single parameter. This allows us to show the validity of our findings regardless of the game's ESS. In the rules every time that an agent $i$ plays, she does so with all her $k_i$ neighbors. An unit of wealth is then distributed among all of them. If everybody cooperates then the payoff is $1/k_i$ for each agent. Otherwise, each defector is given priority and takes a portion $p$ as payoff. If the total amount requested by the defectors, $p \, n_i^D$, is larger than $1$ nobody takes anything. On the contrary, if  $p \, n_i^D \le 1$, the cooperators evenly divide the remaining $1- p \, n_i^D$. Note that for low values of $p$, $p < 1/k_i$, collaboration is the strategy with the largest payoff and in a pure evolutionary framework becomes the only survival. The same occurs on the other extreme for high values of $p$, strictly speaking for $p>1$ defection has a zero payoff. In the area of intermediate $p$ values, the equilibria of our system are equivalent to that of the public goods game and show the effects of the tragedy of commons dilemma
because defection is the most advantageous strategy but if every agent opts for it none of them get any payoff~\cite{hardin68,hardin95}.

After describing the opinion dynamics and the game rules, it is important to explain how both are coupled. As illustrated in Figure~\ref{sketch}, the two extremes of the opinion variable $\phi$ are identified with the strategies $D$ and $C$. $\phi$ represents thus the opinion of the agents about which is the best strategy to win the game. The pass from an agent's opinion  to real action is taken by assuming a probability $p_C = (1+\phi)/2$ of playing $C$ and $p_D = 1-p_C = (1-\phi)/2$ of choosing $D$. It is important to stress that the game is actually played in a mixed strategy framework and that this way of implementing opinion and action is assuming incomplete information, actions based on impressions and a social component in the way the players move towards the selection of a strategy. In practice, the model is updated by choosing a random agent $i$ in each time step, then she plays the game with her neighbors and after this her opinion is updated depending on the earned payoff. For updating the opinions, a neighbor of $i$, $j$, is randomly selected and the new opinions are calculated using Deffuant's model of Eq.~(\ref{deff}) only if $j$'s payoff is equal or higher than $i$'s. Note that only $i$'s opinion is updated, which introduces an asymmetry in Deffuant's rules. This asymmetry prevents players that are doing better from changing opinion due to interactions with other performing worse, and it also breaks the strong conservation of the average opinion that is a feature of the original Deffuant's model.

\section{Results}

\begin{figure}
\begin{center}
\includegraphics[width=8.6cm]{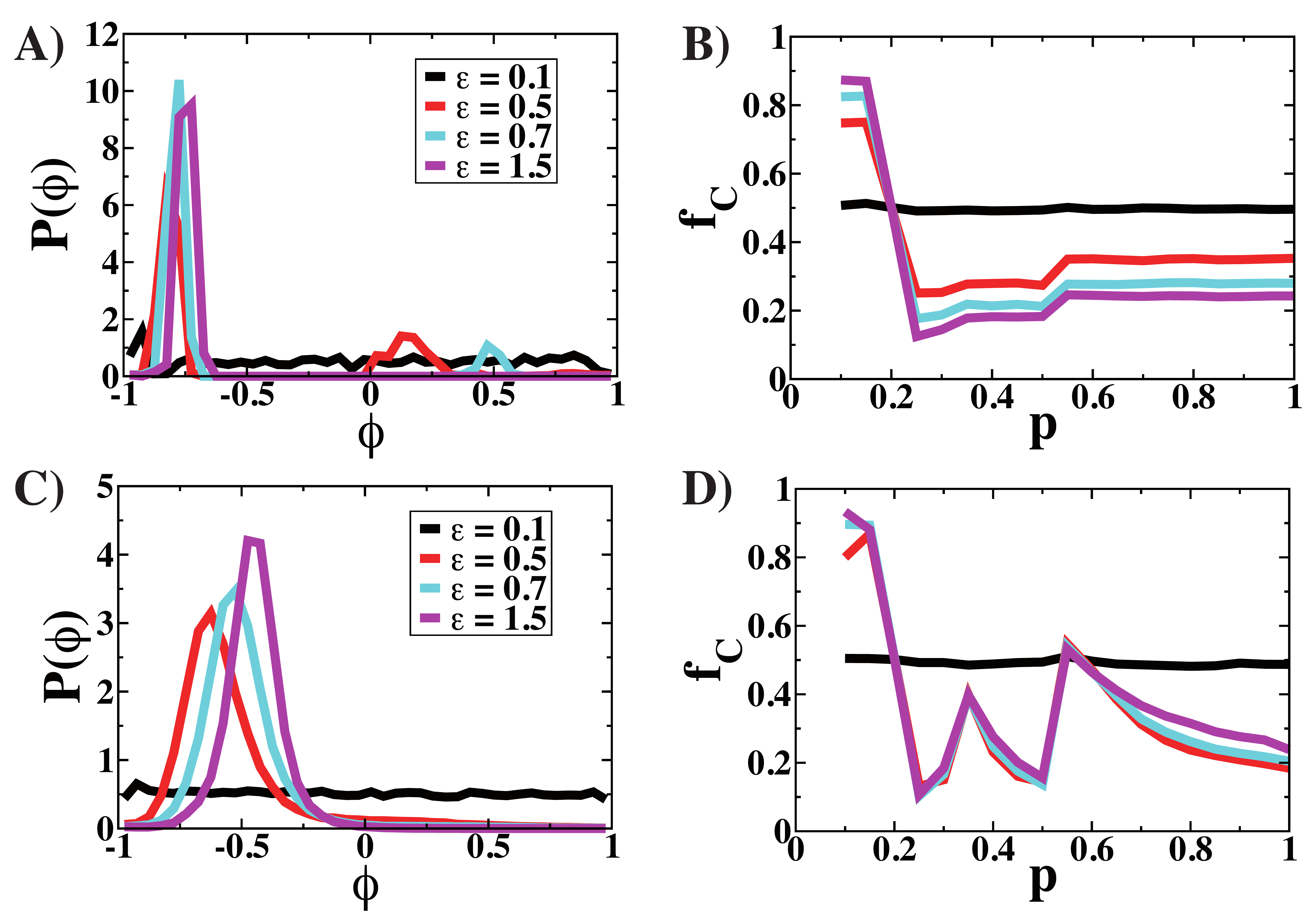}
\caption{Distribution of agent opinions and the average fraction of collaborators $f_C$ as a function of $p$ for different values of the bounding parameter $\epsilon$. In the first plots, A) and B), the opinion update is based on the payoff obtained in the last round of the game, while in C) and D) is  based on the accumulated wealth. In A) and C), $p = 0.8$. The total population in the simulations is $N = 5000$.\label{param}}
\end{center}
\end{figure}

\begin{figure*}
\begin{center}
\includegraphics[width=18cm]{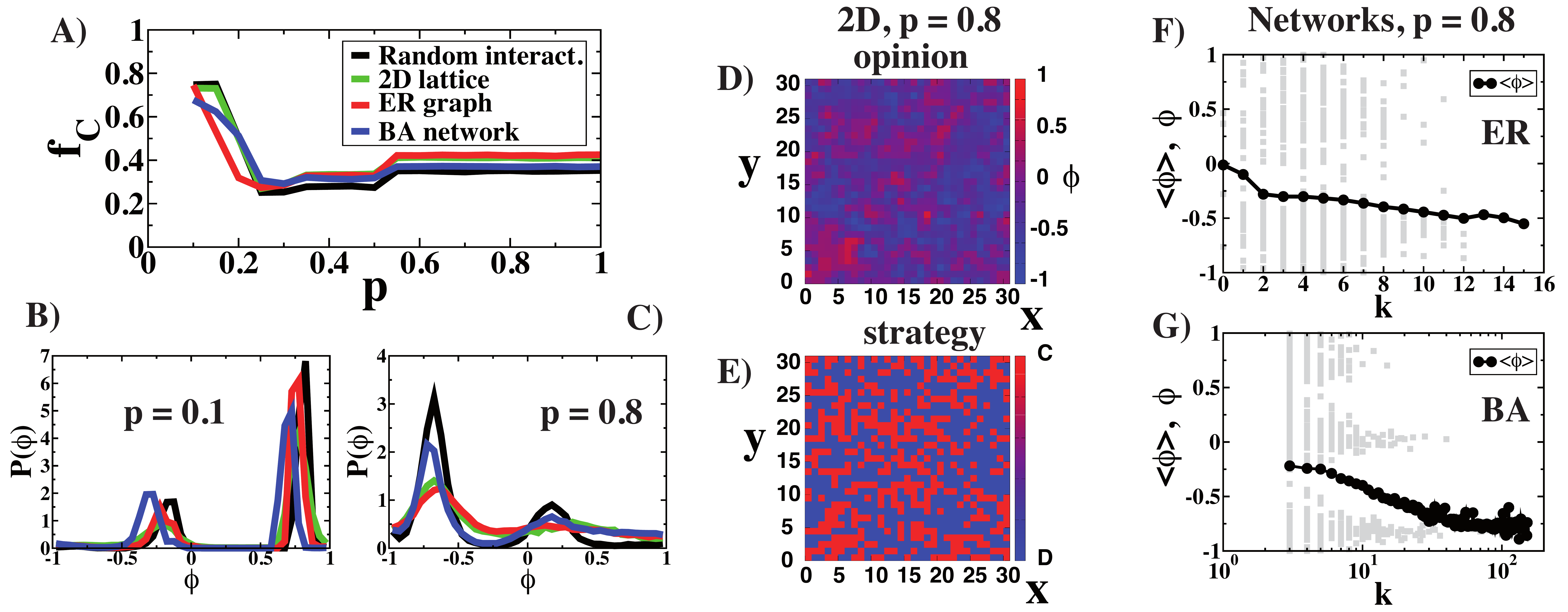}
\caption{Influence of the topology of the interaction network on the outcome of the game. In A), the fraction of collaborators $f_C$ as function of the parameter $p$. In B) and C) the opinion distribution for $p = 0.1$ and $p = 0.8$. Remember that the nature of the game passes from a harmony game to public goods game dilemma at $p = 0.2$. In D) and E), maps showing the opinion and strategy played in an instance of the game. And in F) and $G$, in the background in grey the agents' opinion for a realization for the game and the average opinion for $100$ realizations as a function of the agents' degree $k$. In all cases $\epsilon = 1/2$ and the sizes of the systems are $N = 1000$ for all the systems except the $2D$ lattices that count with $32\times32 = 1024$ agents. The networks are built with $\langle k \rangle \approx 3$.   \label{topol}}
\end{center}
\end{figure*}

Let us start by considering a mean-field situation in which in each time step a randomly selected agent interacts with a group formed by four other agents chosen at random. The first results can be seen in Figure~\ref{comp} where the average opinion $\langle \phi \rangle$ and the average fraction of cooperators $f_C$ are displayed as a function of time. The curves of different colors correspond to three values of $p$: $p = 0.1$, $p = 0.4$ and $p = 0.8$. For games with $5$ participants, cooperation $C$ is the most advantageous strategy below $p < 1/5 = 0.2$. In general if the number of players per game is $n$, the particular value of $p$ for which $C$ is the best strategy is given by $p < 1/n$. Similar results to those desribed next are found for any value of $n$ as long as the values of $p$ are consitently updated. The blue curves ($p = 0.1$) correspond thus to a harmony game, where the $C$ strategy becomes prevailing in the system from an evolutionary perspective. This is actually the outcome when the state of the system is updated following a replicator dynamics (see plots on the left column of Figure~\ref{comp}). Otherwise, for $p > 0.2$, the replicator dynamics results in a final state formed mainly by defectors.  The update based only on the opinion dynamics, without allowing any coupling between the payoff of the game and the update of the opinions, leads to the selection of a few opinion values. These  values of $\phi$ are separated more than $2 \, \epsilon$ and depend on the model initial conditions. The variability of the initial conditions causes the slight dispersion in the distributions $P(\phi)$. This is the known final state for the Deffuant model~\cite{castellano09,deffuant00}. 

More interestingly, the combination of both game and opinion dynamics on the right-hand plots produces a final state that do not correspond with any of the fixed points of the uncoupled dynamics. Although the defectors are still a minority for $p = 0.1$ and a majority for the other values of $p$, the dispersion of opinions is noticeable and a small reservoir of agents with opinion opposite to the majority remains. The origin of this small group of agents lies in the difference between the social and the evolutionary dynamics. Bounded confidence prevents the interaction of agents with very different opinion regardless of their difference in payoff. The members of the small group of roguish agents can play with any other agent but they only update their opinion when confronted with their own peers. This behavior would be eliminated in an evolutionary framework, where the payoff and the fitness are strictly related but this is not necessarily the case in a social environment. Actually, this kind of stubbornness against facts has been observed in behavioral economics where persons are asked to play a repeated Prisoner's dilemma. A fraction of the participants opted for pure defection or even pure collaboration despite the existence of more advantageous strategies such as tip for tap or a Markovian response~\cite{traulsen10,grujic10,gracia12,gracia12b}. These experiments also show a continuous strategy exploration by the participants that may not be so certain of their own choices.

The fact that the small group of contrarian players dissolves when the social constraints are relaxed can be observed in Figure~\ref{param}. In the plot A), the distribution of agents' opinions is displayed for different values of the bound confidence parameter $\epsilon$. If $\epsilon$ is very low there is very few interaction between agents and therefore the opinions remain frozen in the initial condition, which is an uniform distribution. When $\epsilon$ increases, the agents are able to interact with other players holding very different opinions. This leads to the convergence of opinions to values close to the extreme $\phi = -1$, which corresponds to pure defection and that in the dilemma with evolutionary dynamics is the only ESS. The players recognize thus defection as the most adequate strategy in the limit $\epsilon \to 2$ but due to the stochastic nature of the relation between opinions and action are not able to reach $\phi = -1$.  These results are stable within each of the two games to the variation of the values of the portion taken by the defectors $p$. The average fraction of cooperators $f_C$ can be seen in Figure~\ref{param}B as a function of $p$. For all the values of $\epsilon$, a change can be observed in $p = 0.2$ coinciding with the modification of the nature of the game from harmony to a dilemma. Apart from this, some minor corrections are seen due to the discreteness of the group of players. Since only $5$ players are considered in each round and if $n_D$ stands for the number of defectors in a round, the total payoff reserved for the defectors is $n_D \, p$. If this amount goes over the unit neither defectors or collaborators get any payoff. Therefore, the maximum number of defectors that a round can sustain comply with the relation $n_D \, p > 1$. The values of $p$ coinciding with $1/n_D$ mark thus a change on the payoff partition in the game. A final detail that we also wanted to explore here is the stability of the solutions if the total wealth is taken as main factor of the opinion update instead of the instantaneous payoff. The use of the total wealth adds a more consistent memory effect since the choice of a successful strategy allows for a continuous income. Still the players are able to recognize the optimal strategy for large values $\epsilon$, but it is important to note the large dispersion of opinions and the peak around $\phi \approx -0.3$ far from the extreme $\phi = -1$. Also the stability of the system with $p$ becomes altered with more violent bumps in $f_C$ when $p$ passes through the fractional values modifying the payoff partition.

A simplistic mean-field configuration is not a valid match to the more complicated structure that social interactions can present. The social interactions are normally modeled as network whose vertices and edges represent individuals and social relations, respectively. In theoretical works, it has been shown that the topological characteristics of such networks can affect the game outcome increasing, for instance, the level of cooperation in the Prisoner's dilemma~\cite{santos06,szabo07,santos08}. However, experimental results where real individuals play the Prisoner's dilemma with different network topologies contradict this conclusion since the level of cooperation seems to be similar for different network topologies~\cite{traulsen10,grujic10,suri10,gracia12,gracia12b}. The explanation provided for this effect is the presence of the so-called {\em moody conditional cooperators}: individuals that take their strategic decisions regarding cooperation or defection based on their previous experience as much as on their neighbors' payoff. The results of our model point in the same direction with a very weak dependence on the topology of the interaction networks as can be seen in Figure~\ref{topol}. In order to introduce different interaction topologies, we run the model on a $2D$ square lattice, on Erd\"os-Renyi (ER) graphs~\cite{erdos59} and on Barabasi-Albert (BA) scale-free networks~\cite{barabasi99}. The ER and BA graphs are particular types of complex networks with different level of heterogeneity in the number connections of the nodes (degree, $k$). For ER, the distribution of degrees is Poissonian centered around the average $\langle k \rangle$, while for the BA the distribution of degree is a power-law decaying function with exponent $-3$, $P(k) \sim k^{-3}$. In each case, an agent plays each round of the game with her nearest neighbors alone. In Fig.~\ref{topol}A, the fraction of cooperators $f_C$ is displayed as a function of the parameter $p$ for different network topologies and $\epsilon = 1/2$. The fraction of cooperators is not very sensitive to the topology. One can find a slightly stronger difference in the distribution of opinions as can be seen in Figure~\ref{topol}B and C, where it can be seen that a model with random interactions or scale-free networks have more marked peaks. We have also explored the spatial distribution of opinions and strategies when the game is played in a $2D$ square lattice with $4$ neighbors per node (Fig.~\ref{topol}D and E). As occurs with the Prisoner's dilemma in replicator dynamics~\cite{nowak92}, the reduced dimensionality allows for the formation of clusters of agents with close opinions playing similar strategies. The local character of the interactions makes that clusters of collaborators can survive. In Figure~\ref{topol}, we explore also the effect that the heterogeneity in the degree of the agents in the social networks can have on the opinion. The agents' opinion in an instance and the average opinion over many realizations is displayed as a function of the agents degree (plots F and G). The average opinion tends to be more negative, closer to defection, for better connected agents regardless of the particular characteristics of the network.

\begin{figure}
\begin{center}
\includegraphics[width=8.cm]{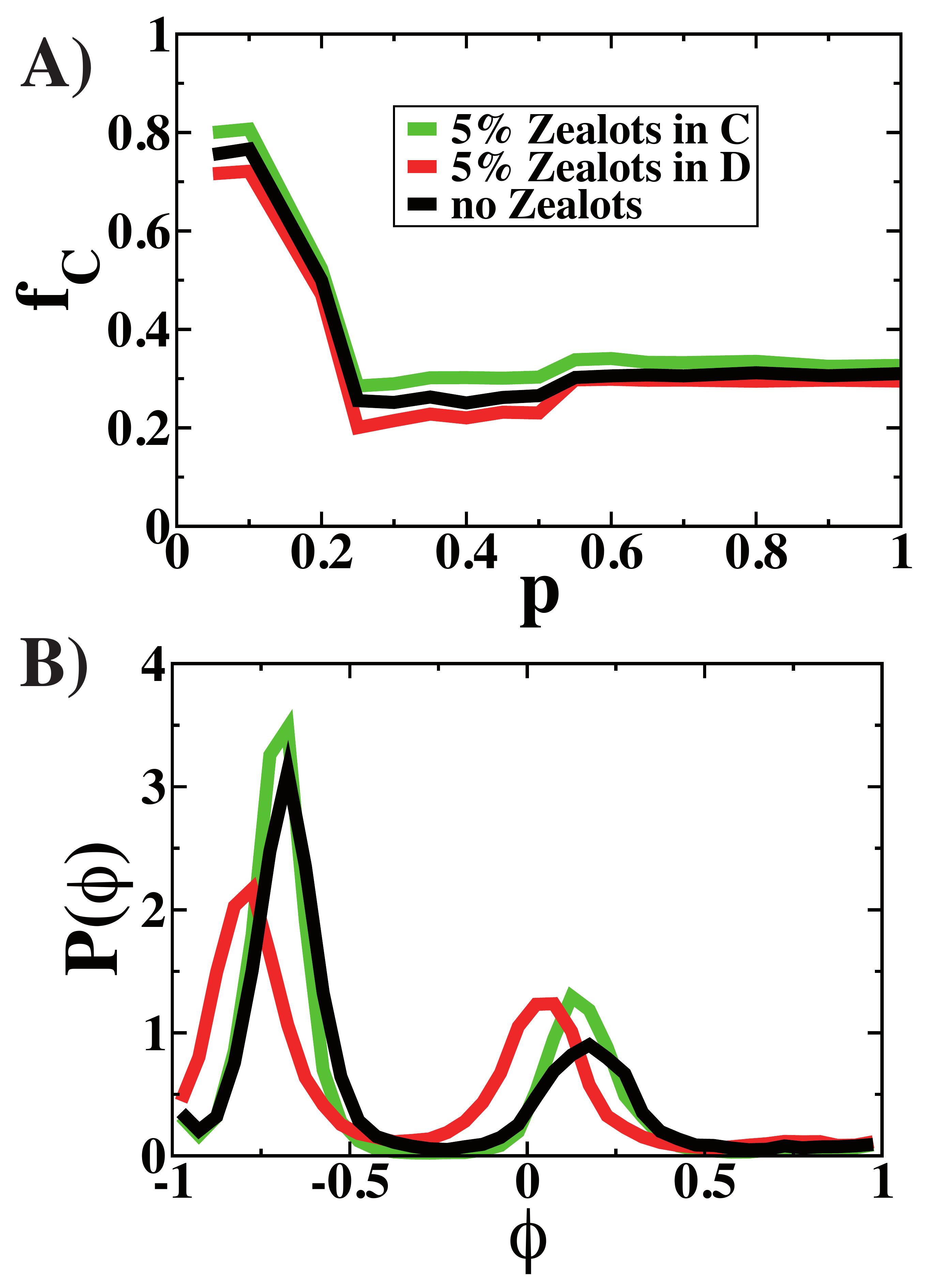}
\caption{Effect of the introduction of zealots on the fraction of collaborators $f_C$ shown in A) as function of $p$ and on the distribution of opinions for $p = 0.8$ in B). The model is simulated for a system of size $N = 1000$, with $\epsilon = 1/2$ and with random interactions.\label{zeal}}
\end{center}
\end{figure}

A final aspect of the model that we analyze is the effect that a small fraction of radical agents can have on the opinion and strategies played by the rest of the population. There are two precedents that justify the concern with the role that the extremists can play. One is the existence of such radical individuals playing always the same strategies either cooperation or defection in the experiments~\cite{gracia12,gracia12b}.  The second is that the effect of extremists, who go under the name of contrarians or zealots in the literature, is well known in the opinion dynamic models~\cite{galam04,galam05}. A small fraction of extremists can drive the system out of consensus. The fraction of cooperators obtained with the model as a function of $p$ and the opinion distribution for $p = 0.8$ are depicted in Figure~\ref{zeal}. The curves for the model with a fraction of extremists of $5\%$ either of players $C$ or $D$ are over-imposed to the baseline without extremists. As can be seen, the average fraction of collaborators $f_C$ is weakly dependent on the presence of extremists or zealots. Apart from a slight shift due to the additional $5\%$ players of pure strategies, no major change is observed. However, the same cannot be said regarding the opinion distributions. Both models with extremists show different distributions even though the effect is more dramatic if the zealots are playing "defect".   

\section{Discussion}  

In summary, we have introduced a model that couples opinion dynamics and strategies selection in a game. Our main assumptions are that the agents have not certainty on which strategy is optimal and that they form an opinion on this issue which can be updated by social pressure. In particular, for the game we have selected a model based on the rules proposed in the Tragedy of Commons by G. Hardin that allows us to explore two possible final equilibria by tuning a single parameter $p$. For $p$ below $0.2$, the rules of our system produce a scenario that reminds a harmony game, while for $p > 0.2$ a social dilemma equivalent to the public goods game is found. For the opinion dynamics, we use Deffuant's model that is characterizes by having a continuous opinion variable $\phi$ and a bounded confidence mechanism embodied by the parameter $\epsilon$. If the opinions of two agents are further away than $\epsilon$, no interaction is possible. We take advantage of the continuous nature of $\phi$ to couple opinions and actions via a mixed strategy model. The two available strategies $C$ and $D$ become thus an action that is taken with certainty only in the limits of opinion $\phi$ $1$ and $-1$, respectively. Any intermediate value of the opinion can be translated into a probability of choosing $C$ or $D$ with the bias towards the closest extreme in $\phi$. 

Once the coupling of opinion and game dynamics is on, the outcome of the game changes.  Of course, the model is stochastic and so a certain amount of dispersion in the main descriptive variables is expected due to the inherent randomness. However, variables such as the average fraction of collaborators or the distribution of opinion reach fixed points in the dynamic different from the de-coupled systems that reflect the constraints that opinion and game payoff put on each other. This effect is  enhanced when the parameter $\epsilon$ is decreased imposing a more strict bounded confidence regime. Cooperation can thus be increased with a more social dynamics for the evolution of the strategies but this is not the only feature that calls for attention in our results. The presence of the variable of opinion allows the system to adapt to different interaction topologies or to the existence of extremist players in a very particular way. In correspondence to the empirical observations, in the coupled model the fraction of cooperators is not altered by the consideration of different topologies or by the introduction of extremists. It is the opinion distribution instead which is modified to absorb the impact of the new conditions. In the experiments, this phenomenon was explained by the presence of {\em moody} players that have into account previous strategies when a new strategic decision was taken. In our model this role is played by the memory effect that the opinion variable provides. In this work, we have selected particularly simple rules for the game and the opinion dynamics. In order to gain further insights  in the decision process of real players more theoretical and experimental work is needed. Nevertheless, the interplay between opinion and actions and the fact that the opinion gets updated by social pressure can significantly modify the scenario in evolutionary games.

\begin{acknowledgments}
FG receives funding from the French Research National Agency through the project SIMPA. JJR is funded by the Ram\'on y Cajal program of the Spanish Ministry of Economy and Competitiveness (MINECO). Partial support was also received from MINECO through the project MODASS (FIS2011-24785).
%and from the European Commission through the project EUNOIA (EC-ICT contract no. $318367$).
\end{acknowledgments}

\end{document}